\documentclass[prx,superscriptaddress,preprint]{revtex4-2}
\usepackage[utf8]{inputenc}
\usepackage[T1]{fontenc}
\usepackage{graphicx}
\usepackage{comment}
\usepackage{amsmath}

\begin{document}
\title{Electronic and structural fingerprints of charge density wave excitations in extreme ultraviolet transient absorption spectroscopy}

\author{Tobias Heinrich}
\affiliation{Max Planck Institute for Multidisciplinary Sciences, 37077 Göttingen, Germany}
\author{Hung-Tzu Chang}
\affiliation{Max Planck Institute for Multidisciplinary Sciences, 37077 Göttingen, Germany}
\affiliation{Department of Chemistry, University of California, Berkeley, California 94720, USA}
\author{Sergey Zayko}
\affiliation{Max Planck Institute for Multidisciplinary Sciences, 37077 Göttingen, Germany}
\author{Kai Rossnagel}
\affiliation{Institute of Experimental and Applied Physics, Kiel University, 24098 Kiel, Germany}
\affiliation{Ruprecht Haensel Laboratory, Deutsches Elektronen-Synchrotron DESY, 22607 Hamburg, Germany}
\author{Murat Sivis}
\affiliation{Max Planck Institute for Multidisciplinary Sciences, 37077 Göttingen, Germany}
\affiliation{4th Physical Institute – Solids and Nanostructures, University of Göttingen, 37077 Göttingen, Germany}
\author{Claus Ropers}
\affiliation{Max Planck Institute for Multidisciplinary Sciences, 37077 Göttingen, Germany}
\affiliation{4th Physical Institute – Solids and Nanostructures, University of Göttingen, 37077 Göttingen, Germany}

\date{\today} 
\begin{abstract}
    Femtosecond core-level transient absorption spectroscopy is utilized to investigate photoinduced dynamics of the charge density wave in 1T-TiSe$_2$ at the Ti M$_{2,3}$ edge (30-50 eV). Photoexcited carriers and phonons are found to primarily induce spectral red-shifts of core-level excitations, and a carrier relaxation time and phonon heating time of approximately $360$ fs and $1.0$ ps are extracted, respectively. Pronounced oscillations in delay-dependent absorption spectra are assigned to coherent excitations of the optical $A_{1g}$ phonon ($6.0$ THz) and the $A_{1g}^*$ charge density wave amplitude mode ($3.3$ THz). By comparing the measured spectra with time-dependent density functional theory simulations, we determine the directions of the momentary atomic displacements of both coherent modes and estimate their amplitudes. 
    This work presents a first look on charge density wave excitations with table-top core-level transient absorption spectroscopy, enabling simultaneous access to electronic and lattice excitation and relaxation.
\end{abstract}
\maketitle

\section{Introduction}
\begin{figure*}
\centering
\includegraphics[width=\textwidth]{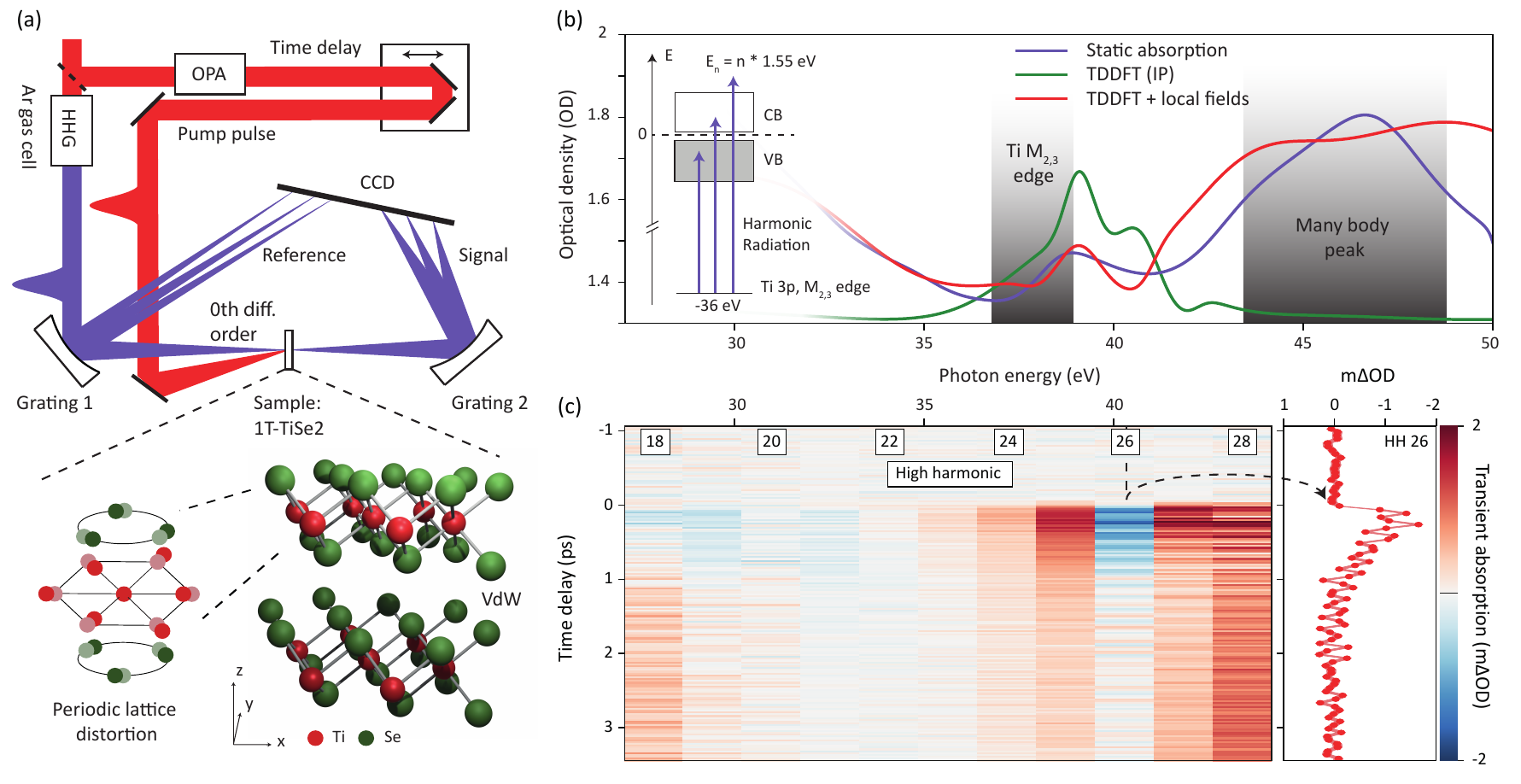}
\caption{(a) Setup for extreme ultraviolet (XUV) transient absorption spectroscopy. Fluctuations of the high-harmonic generation (HHG) source are tracked with a reference spectrum. The 1T-TiSe$_2$ specimen consists of layers held by van-der-Waals (vdW) forces and shows a charge density wave with an associated periodic lattice distortion. (b) Static absorption spectrum recorded with a continuous HHG source \cite{changCoupledValenceCarrier2021}. Density functional theory (DFT) calculations with the inclusion of local field effects resemble the Ti M$_{2,3}$-edge ($ \approx 36.5\, \text{eV}$) as well as the many-body absorption peak at $42 \, \text{eV} - 49\, \text{eV}$. Inset: Titanium M edge originating from transitions of the Ti 3p orbital to empty states in the valence (VB) and conduction band (CB). (c) Exemplary transient absorption spectrum recorded with discrete harmonics spaced by $1.55$ eV. Relative changes to an un-pumped sample are shown for various time delays of the pump pulse.}
\label{fig:Setup}
\end{figure*}

Electronic processes in quantum materials including superconductors, Mott insulators, and charge-density-wave (CDW) compounds encompass a wide range of phenomena with collective excitations involving coupled electronic, vibrational, and spin dynamics \cite{grunerDensityWavesSolids1994, imadaMetalinsulatorTransitions1998,dagottoComplexityStronglyCorrelated2005,basovElectrodynamicsCorrelatedElectron2011}. The layered transition metal dichalcogenide 1T-TiSe$_2$ features CDW formation below a critical temperature ($T_c$) of 200 K \cite{disalvoElectronicPropertiesSuperlattice1976}, linked to a periodic lattice distortion (PLD) in the form of a $2\times 2\times 2$ supercell \cite{wooSuperlatticeFormationTitanium1976}, shown in Fig.~\ref{fig:Setup}(a). Various mechanisms, including Jahn-Teller effects and exciton condensation, were proposed to explain the CDW formation \cite{hughesStructuralDistortionTiSe1977,whangboAnalogiesConceptsMolecular1992,rossnagelChargedensitywavePhaseTransition2002,vanwezelExcitonphonondrivenChargeDensity2010,cercellierEvidenceExcitonicInsulator2007,monneyProbingExcitonCondensate2010,kogarSignaturesExcitonCondensation2017,rossnagelOriginChargedensityWaves2011}. Recent experiments on ultrafast nonthermal melting of the CDW in 1T-TiSe$_2$ suggest that the electronic and vibrational degrees of freedom are strongly coupled in the quenching process \cite{mohr-vorobevaNonthermalMeltingCharge2011,HUBER2022110740,rohwerCollapseLongrangeCharge2011,weberElectronPhononCouplingSoft2011,hellmannTimedomainClassificationChargedensitywave2012,mathiasSelfamplifiedPhotoinducedGap2016,monneyRevealingRoleElectrons2016,karamStronglyCoupledElectron2018,hedayatExcitonicLatticeContributions2019,burianStructuralInvolvementMelting2021,duanOpticalManipulationElectronic2021,hedayatInvestigationNonequilibriumState2021,ottoMechanismsElectronphononCoupling2021} and different timescales of the loss of electronic and structural orders were observed \cite{porerNonthermalSeparationElectronic2014,chengLightinducedDimensionCrossover2022}. 

To understand such complex photophysical phenomena, simultaneous probing of the electronic and lattice subsystems in a single experiment is highly beneficial, as it allows for a study of both degrees of freedom at identical experimental conditions. Core-level transient absorption spectroscopy, in which the sample is pumped with a femtosecond optical pulse and subsequently probed by an extreme ultraviolet (XUV) pulse, is ideally suited for this purpose. This method has been successfully utilized to simultaneously observe the decay of photoexcited carriers and coherent phonons in MoTe$_2$ \cite{attarSimultaneousObservationCarrierSpecific2020} and to disentangle the intricate electron phonon dynamics in graphite \cite{sidiropoulosProbingEnergyConversion2021}. However, up to now, XUV radiation has not been used to study excitations of CDWs. In this work, we apply core-level spectroscopy at the Ti M$_{2,3}$ edge (32-50 eV) to investigate photoinduced excitations of the CDW in 1T-TiSe$_2$. Thereby, we analyze the transient absorption by comparing with \textit{ab initio} simulations conducted with time-dependent density functional theory (TDDFT). We distinguish the electronic and phonon contributions to the core-level transient absorption spectra, extract the timescales of hot electron cooling and phonon heating, separate contributions from different coherent phonon modes by spectral fingerprints, and identify their corresponding atomic displacements, excitation efficiency, and dephasing times.

\section{Results}
The core-level transient absorption experiment is conducted with a 65-nm thick 1T-TiSe$_2$ flake on a 30-nm thick silicon membrane (Appendix~\ref{Appendix Sample}) at a temperature of 110 K. Details of the experimental setup (Fig.~\ref{fig:Setup}(a)) are described in Appendix~\ref{Appendix Setup} and Ref.~\cite{heinrichArxiv}. In brief, the sample is optically excited by a $40$-fs laser pulse centered at 2-$\mu$m wavelength and probed by a time-delayed XUV beam covering the spectral range of 25-50 eV. The XUV radiation is produced through high harmonic generation (HHG) driven by 35-fs long laser pulses (800 nm center wavelength) and their second harmonic in an Ar-filled gas cell. The two-color field creates both even and odd harmonics of the fundamental driving beam centered at $1.55$ eV. The absorption of the XUV radiation in the investigated spectral range is dominated by transitions of the Ti $3p$ electrons into the valence shell (Fig.~\ref{fig:Setup}(b)). The pump-induced change of the XUV absorbance is defined as the difference between the absorbance (optical density, OD) with and without optical excitation ($\Delta \text{OD}=\text{OD}_\text{pumped}-\text{OD}_\text{unpumped}$). An additional reference spectrum of the harmonic source is simultaneously collected for noise suppression, which provides an improved sensitivity beyond $10^{-4}$ OD \cite{heinrichArxiv} and enables the detection of subtle CDW dynamics, as found in 1T-TiSe$_2$ at low pump fluences. More specifically, the investigation of CDW excitations is largely restricted to fluences below the threshold for non-thermal CDW melting \cite{chengLightinducedDimensionCrossover2022}.

Figure ~\ref{fig:Setup}(b) displays a static absorption spectrum of 1T-TiSe$_2$ (purple line), recorded with a spectrally continuous source \cite{changCoupledValenceCarrier2021}. Here, the Ti M$_{2,3}$-edge exhibits a small peak with an onset of approximately 36.5 eV and another strong absorption peak centered at about 47 eV. To understand the Ti M$_{2,3}$-edge transitions, we compare the measured static spectrum with TDDFT simulations (Appendix~\ref{Appendix DFT}) under random phase approximation (RPA). The simulated transitions within the independent particle (IP) approximation (Fig.~\ref{fig:Setup}(b), green line) only overlaps with the $\sim$39 eV small peak in the empirical spectrum, whereas the strong peak at $\sim$47 eV is absent. Calculations including local field effects (Fig.~\ref{fig:Setup}(b), red line) \cite{krasovskiiLocalFieldEffects1999}, where many-body interactions in electronic excitations are partially accounted for, qualitatively reproduce the empirical spectrum. This suggests that the small peak at the onset of the Ti M$_{2,3}$-edge (36.5 eV) mainly constitute single-particle excitations from the Ti $3p$ orbitals to the conduction band. In addition, the strong peak at $\sim$47 eV can be identified as a giant resonance \cite{conneradeGiantResonancesAtoms1987,amusiaTheoryCollectiveMotion2000} comprising transitions from the Ti $3p$ levels to the conduction band, accompanied by valence-shell excitations from below to above the Fermi level through configuration interactions. Similar features are also observed in the Ti M$_{2,3}$ edge absorption in elemental titanium \cite{volkovAttosecondScreeningDynamics2019}. 

The photoinduced dynamics in 1T-TiSe$_2$ are tracked by recording XUV absorption spectra as a function of the pump-probe time delay $\tau$ (Fig.~\ref{fig:Setup}(c)). Due to the discrete spectrum of the harmonic source (Appendix~\ref{Appendix Setup}) and the weak pump-induced signals ($<10^{-3}$ OD), the spectra are binned to the center of each harmonic peak, separated by the photon energy of the fundamental driving laser (1.55 eV). At pump-probe overlap ($\tau\approx 0$), the absorbance decreases at energies below 35 eV. An increase in absorbance is observed in both the Ti M$_{2,3}$ near-edge region (35-39 eV) and at the onset of the many-body giant resonance (42-45 eV). In-between the tail of the lower-energy peak of Ti M$_{2,3}$ edge (39 eV) and the onset of the giant resonance (42 eV), a decrease in absorbance is found. At long time delays ($\tau >3$ ps), an absorbance increase is seen across the entire spectrum. At time delays up to $\sim$3 ps, oscillations in transient absorption signals are evident from the 26$^{th}$ to the 31$^{th}$ harmonic (Fig.~\ref{fig:Setup}(c) right). In the following sections, we analyze the transient absorption signals at pump-probe overlap and $>3$ ps time delay (Sec.~\ref{Sec:HotElectron}), discuss the origin of the oscillations (Sec.~\ref{Sec:Phonon}), and extract the timescales of the underlying physical processes.

\subsection{Hot electron dynamics and lattice heating}
\label{Sec:HotElectron}
\begin{figure*}
\centering
\includegraphics[width=\textwidth]{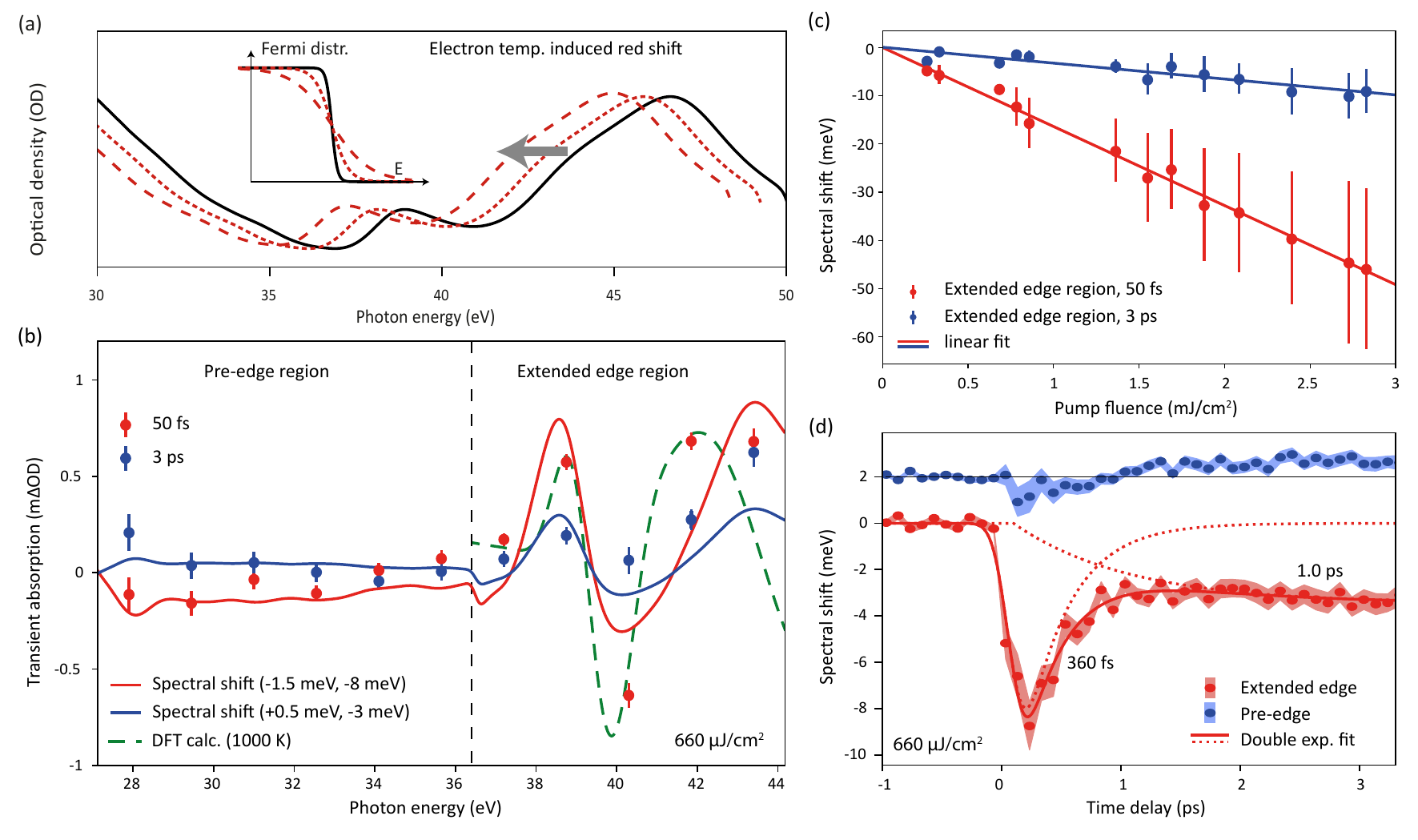}
\caption{(a) Schematic spectral changes induced by electronic heating. (b) Transient absorption spectra. The energy-dependent response can be modeled with selective spectral shifts of the static spectrum (Fig. 1 (b)). To this end, the spectral range is separated into the pre-edge  ($< 36.5 \, \text{eV}$) and extended edge ($> 36.5 \, \text{eV}$) regions. Complementary DFT calculation for elevated electron temperature (green dashed line) shows a similar spectral change. (c) Fluence dependence of the extended edge shifts showing a linear scaling in support of thermal mechanisms. (d) Temporal dynamics of pre-edge and extended edge region spectral shifts. A double-exponential fit extracts electronic excitation decaying on $\approx 360 \, \text{fs}$ and contributions of the phonon bath that emerge on the ps time scale.}
\label{fig:spectralShift}
\end{figure*}

Transient absorption signals at $\tau=50$ fs and 3 ps are plotted in Fig.~\ref{fig:spectralShift}(b) (dots). As the energy transfer from the electrons to the lattice typically occurs on the hundreds-of-femtoseconds to picosecond timescale \cite{hedayatExcitonicLatticeContributions2019,ottoMechanismsElectronphononCoupling2021}, photoexcited carriers are expected to be the main contributor to the transient absorption signal at 50-fs delay. In many materials, photoexcited carriers contribute to a positive core-level transient absorption signal below the edge and a negative signal above the edge due to holes and electrons below and above the Fermi level, respectively \cite{zurchDirectSimultaneousObservation2017,zurchUltrafastCarrierThermalization2017,schlaepferAttosecondOpticalfieldenhancedCarrier2018,linCarrierSpecificFemtosecondXUV2017,attarSimultaneousObservationCarrierSpecific2020,verkampCarrierSpecificHotPhonon2021}. However, at the onset of the Ti M$_{2,3}$ edge, the transient absorption signal exhibits an increase with energy, contrary to the expected signal contribution from electronic state blocking. This can be explained by significant many-body effects affecting the core-level transitions \cite{volkovAttosecondScreeningDynamics2019,changElectronThermalizationRelaxation2021,changCoupledValenceCarrier2021}. In core-level transient absorption experiments on nickel, it was shown that the transient absorption spectra can be described by a shift and broadening of the static absorption spectrum \cite{changElectronThermalizationRelaxation2021}. Here, we use a similar approach to analyze the transient absorption spectra. First, the spectra are separated into two regions: the region containing the Ti M$_{2,3}$ edge transitions (>36.5 eV), and the region mainly contributed by the tail of the bulk plasmon excitation (<36.5 eV, cf. Fig. 1(b)). By applying red shifts of 1.5 meV and 8 meV to the static absorption spectrum below and above 36.5 eV, respectively, the resulting spectrum shows reasonable agreement with the empirical transient absorption data (Fig.~\ref{fig:spectralShift}(b) red line and dots). 

We compare the measured transient absorption spectrum at 50 fs with TDDFT simulations to explore whether a hot (quasi-thermalized) electronic population is consistent with the observed absorption changes. Figure~\ref{fig:spectralShift}(b) (green dashed line) shows the difference between the core-level absorption spectrum simulated for an electronic temperature at 1000 K and room temperature. 
The simulation shows good agreement with the experimental spectrum, suggesting that at 50-fs an excited electronic population leads to a red shift qualitatively shown in Fig.~\ref{fig:spectralShift}(a). Notably, the DFT calculation is in even better agreement with the data compared to the simplified spectral shift model.

In contrast to the dominance of photoexcited electrons in the transient absorption spectrum at 50-fs time delay, the spectrum at 3-ps delay is expected to be dominated by the heated phonon bath. Nevertheless, the empirical transient absorption spectrum at 3 ps can still be approximated by an energy-shifted static absorption spectrum. Figure~\ref{fig:spectralShift}(b) (blue line and dots) shows that the transient absorption below 36.5 eV has diminished to almost zero, while the spectrum above 36.5 eV can be qualitatively described by a 3 meV red-shifted static absorption spectrum. The agreement between the empirical data and energy-shifted static spectrum degrades in the >40-eV region. This may be explained by the different response to photoexcitations between the Ti M$_{2,3}$ edge onset, which is mainly due to one-body core-level excitations, and the giant resonance at >42 eV, where many-body effects dominate. The absorption in the >40 eV region is thus expected to be influenced by both mechanisms. As the fitting only incorporates one spectral shift parameter, the behaviour of the transient absorption signal cannot be properly described here. 

To relate the spectral shifts to physical observables, we conduct fluence-dependent measurements at these two time delays. The resulting spectral shifts from fitted spectra at >36.5 eV depend linearly on pump fluence (Fig.~\ref{fig:spectralShift}(c)), indicating that the spectral shifts also depend linearly on the carrier and phonon populations. Complementary TDDFT calculations at different electron temperatures confirm a linear dependence of the spectral shift on the electron temperature in the relevant fluence regime. In photoemission studies that offer a more detailed view on the electronic structure, it was shown that electron-electron scattering leads to a quasi-equilibrium within $200$ fs \cite{mathiasSelfamplifiedPhotoinducedGap2016}. In comparison, the spectral redshift is expected to be less sensitive to the finer details of the carrier distribution but serves as a good measure for the effective strength of electronic excitation, as indicated by the linear fluence dependence and the TDDFT calculations. Hence, by extracting the spectral shifts at each time delay, the time-dependent electron and phonon population can be qualitatively tracked. Figure~\ref{fig:spectralShift}(d) (red line) displays the fitted spectral shifts of the extended edge region as a function of time delay. As the shifts comprise contributions from both electrons and phonons, a double exponential fitting is applied. The resulting fitting yields a $360 \pm 40$ fs decay and a $1.0 \pm 0.3$ ps rise, which we assign as the electronic population decay time and phonon population rise time. 

\subsection{Coherent Phonon dynamics}
\label{Sec:Phonon}
\begin{figure*}
\centering
\includegraphics[width=\textwidth]{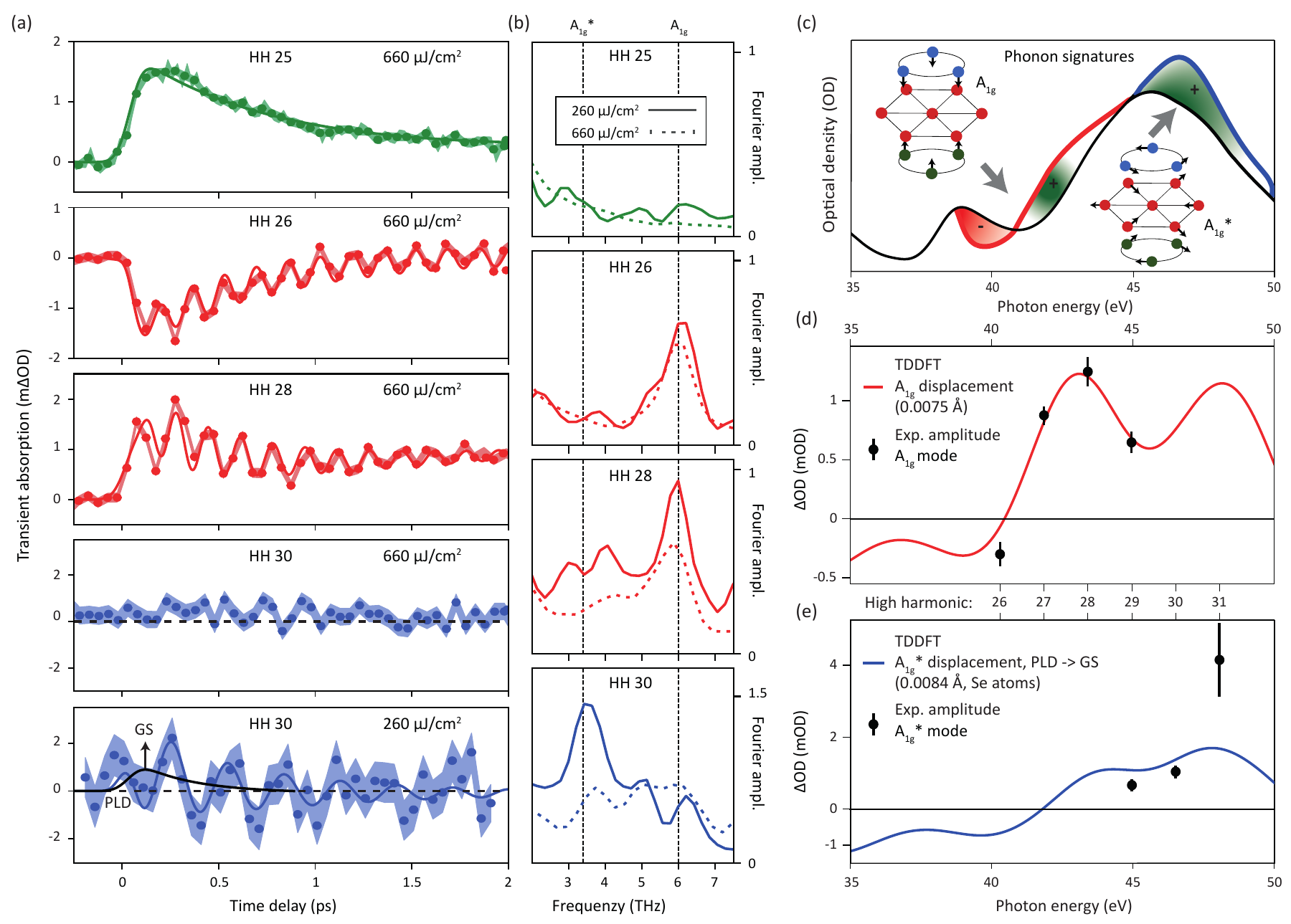}
\caption{(a) Selected transient absorption traces of 1T-TiSe$_2$. Pump-probe traces have been recorded for $F = 260 \, \mu \text{J}/\text{cm}^2$ and $F = 660 \, \mu \text{J}/\text{cm}^2$ at $110 \, \text{K}$ in the CDW phase. (b) Fourier spectra of pump-probe traces. The observed $6 \, \text{THz}$ and $3 \, \text{THz}$ oscillations are attributed to the ${A_{1g}}$ optical mode and the ${A_{1g}}^*$ amplitude mode of 1T-TiSe$_2$. The optical mode is largely independent of the pump fluence while the signature of the amplitude mode is suppressed at large excitation. (c) Spectral fingerprint of the phonon modes. Each mode can be assigned to a characteristic change in the spectrum. (d) Spectral change of the ${A_{1g}}$ displacement calculated by TDDFT compared to the initial oscillation amplitude in the pump-probe traces. (e) Same as (d) but for the ${A_{1g}}^*$ amplitude mode.}
\label{Fig:Phonon}
\end{figure*}
The XUV transient absorption spectra exhibit oscillations as a function of time delay (Fig.~\ref{fig:Setup}(c) right side) that stem from coherently excited lattice modes. In 1T-TiSe$_2$, the harmonic orders 26 to 31 display oscillations at $3.3 \, \text{THz}$ and $6.0 \, \text{THz}$ as shown in Fig.~\ref{Fig:Phonon}(a)\&(b), Fig.~\ref{Fig:PPTraces}, and Fig.~\ref{Fig:Fourier}. From phonon frequencies identified in previous Raman and optical pump-probe studies \cite{hedayatInvestigationNonequilibriumState2021,holyRamanInfraredStudies1977}, the 6 THz and 3.3 THz oscillations are assigned to the $A_{1g}$ optical mode and the $A_{1g}^*$ CDW amplitude mode, respectively. 

The optical mode consists of purely out-of-plane oscillations of all selenium atoms. In contrast, the amplitude mode, which involves the coherent oscillation of the PLD, comprises in-plane movements of both Ti and Se atoms. For comparison, the eigenvectors of these modes are depicted in Fig.~\ref{Fig:Phonon}(c). As the amplitude mode is only found in the PLD-ordered state below the critical temperature \cite{holyRamanInfraredStudies1977}, a loss of its signal was interpreted as a partial suppression of the PLD order \cite{hedayatInvestigationNonequilibriumState2021}. In this study, we observe a corresponding suppression of the amplitude mode by increasing the pump fluence from $F = 260 \, \mu \text{J}/\text{cm}^2$ to $F = 660 \, \mu \text{J}/\text{cm}^2$. At the higher fluence, the 3.3 THz signature is no longer discernible (HH 30, Figs.~\ref{Fig:Phonon}(a)\&(b)), whereas the optical mode amplitude is largely unaffected. 

Interestingly, the spectral changes induced by the two phonon modes are energy-separated and these spectral fingerprints are schematically drawn in Fig.~\ref{Fig:Phonon}(c). From the lineouts shown in Fig.~\ref{Fig:Phonon}(a) and Fig.~\ref{Fig:PPTraces} as well as their Fourier components displayed in Fig.~\ref{Fig:Phonon}(b) and Fig.~\ref{Fig:Fourier}, we identify 3 spectral regions with different contributions to the transient absorption signal. First, at photon energies below 38.7 eV (HH 25, green), no coherent oscillations are visible. Between 40.3-45 eV (red), the transient absorption spectra are dominated by the 6 THz oscillations assigned to the $A_{1g}$ optical mode. Finally, the 3.3 THz oscillations corresponding to the amplitude mode only appears above 45 eV (blue).

The spectral fingerprints of the observed oscillations are further analysed by direct comparison with TDDFT calculations. Here, the spectra are computed for small atomic displacements along the phonon eigenmodes predicted by DFT simulations \cite{biancoElectronicVibrationalProperties2015}, which are schematically shown as black arrows in Fig.~\ref{Fig:Phonon}(c). By comparing the spectra to the absorption of an undistorted lattice, the expected change in the optical density is calculated (see Appendix \ref{Appendix DFT} for more information). In Figs.~\ref{Fig:Phonon}(d) and ~\ref{Fig:Phonon}(e), the calculated OD change is compared to the experimentally observed oscillation amplitude near time zero. The spectral fingerprints are well reproduced by the DFT model. The simulated $\Delta \text{OD}$ for the amplitude mode shows a maximum amplitude at the 31$^\text{st}$ harmonic and a smaller response at the 30$^\text{th}$ and 29$\text{th}$ harmonic. A very different spectral response at harmonics 26 to 29 is observed for the optical mode. Here, the sign reversal between harmonics 26 and 27, which can also be seen by comparing HH 26 and HH 28 in Fig.~\ref{Fig:Phonon}(a), is faithfully reproduced by the calculations.  

With the link to real-space coordinates provided by the TDDFT calculations, the direction of the atomic movements related to positive or negative absorption changes can unambiguously be identified. For the amplitude mode positive absorption changes correspond to a atomic movement from the PLD towards the undistorted ground state (GS). In case of the optical mode, a positive transient absorption of the of harmonic 27 to 29 is indicative of a contraction movement like shown in Fig.~\ref{Fig:Phonon}(c).

Besides the qualitative agreement, a comparison of TDDFT data to the experiment allows us to quantify the initial amplitude of the coherent oscillations. For the optical mode, the experimental absorption change is reproduced by Se atoms displaced by $0.0075 \textup{~\AA}$, and for the amplitude mode by reducing the PLD by 30\% ($0.0084 \textup{~\AA}$ for Se atoms and $0.0255 \textup{~\AA}$ for Ti atoms). The energy stored in the modes can thus be estimated from the real-space displacements and the phonon frequencies. Under the assumption of a homogeneous excitation over the entire sample region probed, stored energies per normal-phase unit cell (u.c.) of $0.65 \, \text{meV}/\text{u.c.}$ and $0.8 \, \text{meV}/\text{u.c.}$ are obtained for the optical and amplitude mode near time-zero, respectively (Appendix~\ref{Appendix Fluence}). These values represent a significant fraction of the $260 \, \mu \text{J/cm}^2$ pump fluence, which translates to an absorbed energy of $6 \, \text{meV}/\text{u.c.}$ (Appendix \ref{Appendix Fluence}). Interestingly, at the larger fluence of $660 \, \mu \text{J/cm}^2$ the amplitude of the optical mode excitation remains largely unchanged compared to the low-fluence data, implying that the efficiency of phonon excitation is nonlinear. This will be further discussed in Sec. \ref{Sec:Discussion}.

In order to obtain meaningful damping times of the coherent modes, the effects from electronic and lattice heating on the pump probe traces need to be separated form the coherent modulation. To this end, we employed a fitting model that incorporates all effects at once (equation \ref{eq:fitting}). Resulting fits are shown as solid lines in Figs. \ref{Fig:Phonon}(a) and \ref{Fig:PPTraces}. Thereby, the seemingly stark difference between the strong damping of harmonic 28 and the weaker damping of harmonic 26 shown in Fig.~\ref{Fig:Phonon}(a) for the same (optical) mode can be explained by the interplay between the characteristic phonon signature and the red shift caused by electronic heating. This effect is included by a cross-term between the phonon and electronic signatures. At fixed probe energy, the cross-term can diminish or enhance the signal of coherent phonons depending on the slope of its spectral fingerprint. For the optical mode, phonon damping constants of $1.1 \pm 0.8$ ps and $1.9 \pm 1.2$ ps are extracted for harmonic 26 and 28 at $660 \, \mu \text{J/cm}^2$. As a reference, the damping without the cross-term is fitted to be $2.3 \pm 0.5$ ps and $0.9 \pm 0.2$ ps for the same data set. The introduction of the cross-term introduces an additional fitting parameter and therefore decreasing the precision of the fit, but reduces the spread of the extracted damping times to an interval within the predicted error. At the lower fluence of $260 \, \mu \text{J/cm}^2$, the agreement is even better. The amplitude mode damping time is analysed at harmonic 30, which shows the best signal-to-noise ratio. For this mode, we find a damping on the $1.3 \pm 1.2$ ps timescale. Despite the large uncertainties, the extracted dephasing time constants of both phonon modes are in good agreement with previous studies \cite{hedayatExcitonicLatticeContributions2019}. 

While harmonic 30 does not show any contribution of an elevated electron temperature at the larger fluence, an additional weak contribution decaying exponentially on the $300 \pm 300 \, \text{fs}$ timescale is found at the lower fluence (black line Fig.~\ref{Fig:Phonon}(a)). We assign this component to an additional displaced atomic position. As discussed earlier, the short-lived positive absorption change can be associated with a mean displacement of the amplitude mode from the PLD state towards the normal phase ground state (GS). A similar behaviour is visible at harmonic 29 in Fig.~\ref{Fig:PPTraces}. Such transient oscillation around a new equilibrium position may be associated to a displacive excitation (DE) mechanism that is expected for the amplitude mode in 1T-TiSe$_2$ \cite{burianStructuralInvolvementMelting2021,monneyRevealingRoleElectrons2016}. However, usually the oscillation of DE occurs around a new equilibrium position and relaxes on larger time scales.

\section{Discussion}
\label{Sec:Discussion}
In this study, we distinguish the suppression of the amplitude mode in 1T-TiSe$_2$ above a certain threshold fluence from the simultaneously excited optical mode. For better comparison to previous experiments, the incident fluences used in this work ($F = 260 \, \mu \text{J}/\text{cm}^2$ and $F = 660 \, \mu \text{J}/\text{cm}^2$) are converted to deposited energy per unit cell (u.c.) (see Appendix \ref{Appendix Fluence}) resulting in $6 \, \text{meV}/\text{u.c.}$ and $15\, \text{meV}/\text{u.c.}$. At the lower excitation density of $6 \, \text{meV}/\text{u.c.}$, the amplitude mode is well visible and suppressed at $15\, \text{meV}/\text{u.c.}$ deposited energy. Considering the strak differences in the experimental setup, sample, temperature and pump wavelength, this threshold is still in reasonable agreement with the previously reported threshold value of $4 \, \text{meV}/\text{u.c.}$ \cite{hedayatExcitonicLatticeContributions2019}. As the amplitude mode only exists in the PLD-ordered state, such a suppression is a clear indication of a loss of PLD order. The expected temperature increase is found to be $\sim 30$ K for the large pump fluence in our experiment which is insufficient to drive the system above the transition temperature of 200 K. Various mechanisms have been proposed to describe this non-thermal melting of 1T-TiSe$_2$, ranging from exciton breaking to a loss of 3D coherence \cite{burianStructuralInvolvementMelting2021,mohr-vorobevaNonthermalMeltingCharge2011,chengLightinducedDimensionCrossover2022,porerNonthermalSeparationElectronic2014}. 

To gain more insights into the mechanism of CDW excitations in 1T-TiSe$_2$, understanding the photo-induced dynamics and the interplay of the electronic and phononic systems is of great importance. We find that the signal from the optical mode persists independent of the fluence, and we do not observe any direct involvement of the optical mode on the quenching of the CDW/PLD state. The observed nonlinear dependence of the excitation efficiency on the fluence is consistent with previous work. A sub-linear scaling has been reported above the threshold \cite{hedayatExcitonicLatticeContributions2019} and our work extends this observation to lower fluences. A possible mechanism could be the indirect influence of the non linearly excited electronic system which determines the excitation efficiency of the phonon mode. While the optical mode's spectral fingerprint resides near the absorption edge, the spectral region of the amplitude mode coincides with the many-body absorption peak. This suggests a strong coupling of the amplitude mode to a perturbation of the electron screening and cooperative electron motion \cite{amusiaTheoryCollectiveMotion2000,ankudinovDynamicScreeningEffects2003} which supports an interplay of exciton condensation and Jahn-Teller mechanism for the formation of a CDW with associated PLD in 1T-TiSe$_2$ \cite{hughesStructuralDistortionTiSe1977,whangboAnalogiesConceptsMolecular1992,rossnagelChargedensitywavePhaseTransition2002,vanwezelExcitonphonondrivenChargeDensity2010,cercellierEvidenceExcitonicInsulator2007,monneyProbingExcitonCondensate2010,kogarSignaturesExcitonCondensation2017,rossnagelOriginChargedensityWaves2011}. A perturbed electron-electron correlation may therefore be one cause of the amplitude mode excitation. Additionally, we find indications of a displacive phonon excitation. It describes the mechanism of an instantaneous change in the atomic potential landscape after electronic excitation, which leads to a motion towards a new equilibrium structure. Indeed, we observe that the optically excited electronic system drives the atoms towards the undistorted ground state. When the electronic system relaxes back, the energy landscape and the equilibrium position of the oscillation follow. The extracted relaxation time of the displaced equilibrium position on roughly $300 \pm 300 \, \text{fs}$ matches the fast recovery of the valence band shift that is usually interpreted as an indicator of the CDW order in 1T-TiSe$_2$ \cite{huberRevealingOrderParameter2022,hedayatExcitonicLatticeContributions2019}. However, such a fast relaxation is unusual for displacive excitation and may hint at a more complex excitation pathway that warrants further investigation. 

At larger timescales in the ps regime, energy transfer between the phonon modes drastically influence the dynamics. Both the optical mode and the amplitude mode are strongly excited and transiently store a significant fraction of the excitation energy, which is redistributed on their respective damping timescales. We find that 11\% of the $6 \, \text{meV}/\text{u.c.}$ excitation ($F = 260 \, \mu \text{J}/\text{cm}^2$) are initially deposited in the optical mode while 14\% are transferred to the amplitude mode. The effect of transient energy storage in the phonon modes can be seen when comparing the extracted timescales. The lattice heats up significantly slower on the $1$ ps time scale compared to the cooling of the electronic system, which occurs on roughly $360$ fs. Such a discrepancy may be explained by the redistribution of additional energy from the $A_{1g}$ optical and the ${A_{1g}}^*$ amplitude mode to other phonon modes within the $\sim 1.3\,\text{ps}$ damping time. Exceptionally large phonon populations can drastically influence the relaxation dynamics and lead to bottleneck effects in 1T-TiSe$_2$ \cite{hedayatExcitonicLatticeContributions2019,storeckStructuralDynamicsIncommensurate2020}, for example, which hinder the reestablishment of the CDW phase.

\section{Conclusion}
In this work, we use high sensitive transient absorption spectroscopy with a HHG source to investigate and distinguish various electronic and lattice contributions of the weakly excited CDW in 1T-TiSe$_2$. Thereby, we paid particular attention to the coherently excited phonon modes. The out-of-plane $A_{1g}$ optical phonon can be separated from the ${A_{1g}}^*$ amplitude mode, as each mode hosts a specific spectral fingerprint. We find, that the amplitude mode is rooted in the many-body absorption peak of the static absorption spectrum, which strengthens the model of a close link between electron-electron correlations and PLD stability. Using TDDFT, the spectral fingerprints can be reproduced and linked to real-space lattice displacements. With the knowledge of displacements and frequencies, the deposited energies and hence the excitation efficiencies of the two modes are estimated. Knowledge of how efficiently the two modes and especially the amplitude mode are excited, is a key aspect of understanding CDW stability as it dictates relaxation dynamics such as the previously observed bottleneck effect \cite{hedayatExcitonicLatticeContributions2019}. Complementary analysis of spectral shifts allows for an extraction of the time scales of electronic cooling and phonon-bath heating on the same data set. Comparing the dynamics of coherent phonons with electronic and lattice temperature helps to form a more comprehensive picture of the CDW quenching.

This work exemplifies the potential of high-sensitivity transient absorption spectroscopy for the investigation of quantum materials by simultaneously probing electronic and lattice degrees of freedom. In future experiments, a combination with spectrally-continuous sources will enable simultaneous lattice and electronic relaxation probing at attosecond timescales. The extracted spectral fingerprints of the phonon modes may prove valuable in further experiments such as imaging with high harmonic sources. In addition, the presented approach may be applied to a variety of phenomena with strongly coupled dynamics of electronic and lattice systems like CDW formation, metal-to-insulator transitions and superconductivity. 

\begin{acknowledgments}
This work was funded with resources from the Gottfried Wilhelm Leibniz Prize and the Deutsche Forschungsgesellschaft (DFG). We thank Stephen R. Leone for providing lab instruments to record continuous XUV spectra of 1T-TiSe$_2$, which was funded by the Air Force Office of Scientific Research (FA9550-19-1-0314 and FA9550-20-1-0334). H.-T. C. acknowledges support from Air Force Office of Scientific Research (FA9550-19-1-0314 and FA9550-20-1-0334). The simulations were conducted at the Scientific Compute Cluster at GWDG, the joint data center of Max Planck Society for the Advancement of Science (MPG) and University of Göttingen.
\end{acknowledgments}

\appendix
\section{Sample preparation}
\label{Appendix Sample}
\begin{figure*}[h!]
\centering
\includegraphics[width=\textwidth]{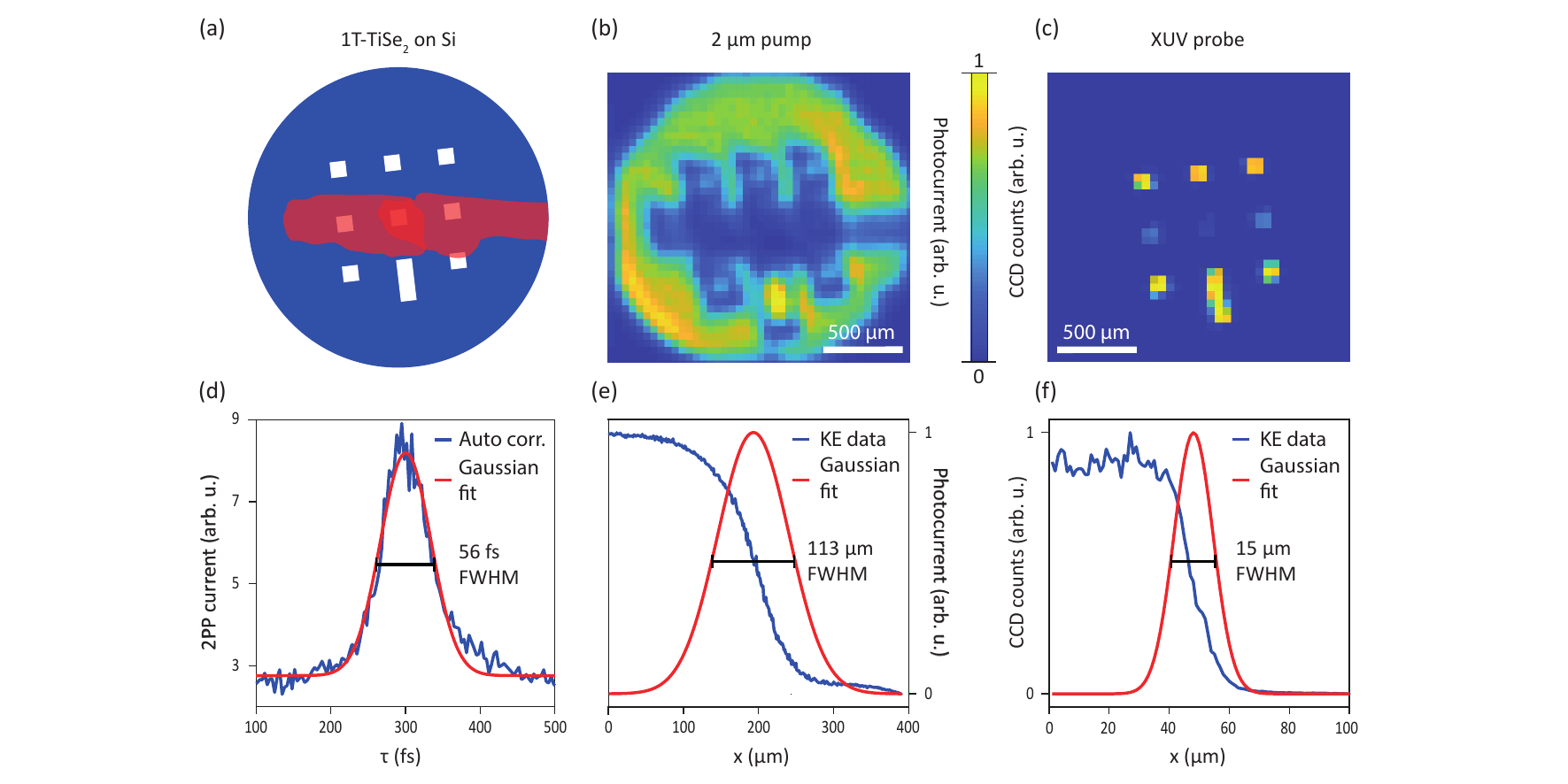}
\caption{(a) Schematic of 9 window Si membrane apertures and a 1T-TiSe$_2$ flake positioned on top. (b) \& (c) Transmitted intensity of pump and probe beam used for spatial overlapping of both beams. (d) Intensity autocorrelation of the pump pulse recorded by 2-photon absorption. (e) \& (f) Knife-edge (KE) measurements of pump and probe beam and estimated Gaussian beam profile.}
\label{Fig:Sample}
\end{figure*}

This study uses 1T-TiSe$_2$ flakes grown by chemical vapor transport \citep{wangControlledSynthesisTwoDimensional2016} and cut to $L= 65\, \text{nm}$ thickness at lateral sizes of $ \approx 300 \, \mu \text{m}$ by ultramicrotomy. The flakes were positioned on a commercial TEM membrane of $30 \, \text{nm}$ thick nanocrystalline, porous silicon (SiMPore) consisting of eight $100\,  \text{x} \,100 \, \mu \text{m}$ and one $ 350 \, \mu \text{m}\, \text{x} \,100 \, \mu \text{m}$ windows. By aligning the sample with the windows, some were fully covered while others remained empty to allow for absolute transmission measurements compared to the pure silicon transmission. A schematic sample is depicted in Fig.~\ref{Fig:Sample}(a).

\section{Transient absorption setup}
\label{Appendix Setup}
\begin{figure*}[h!]
\centering
\includegraphics[width=\textwidth]{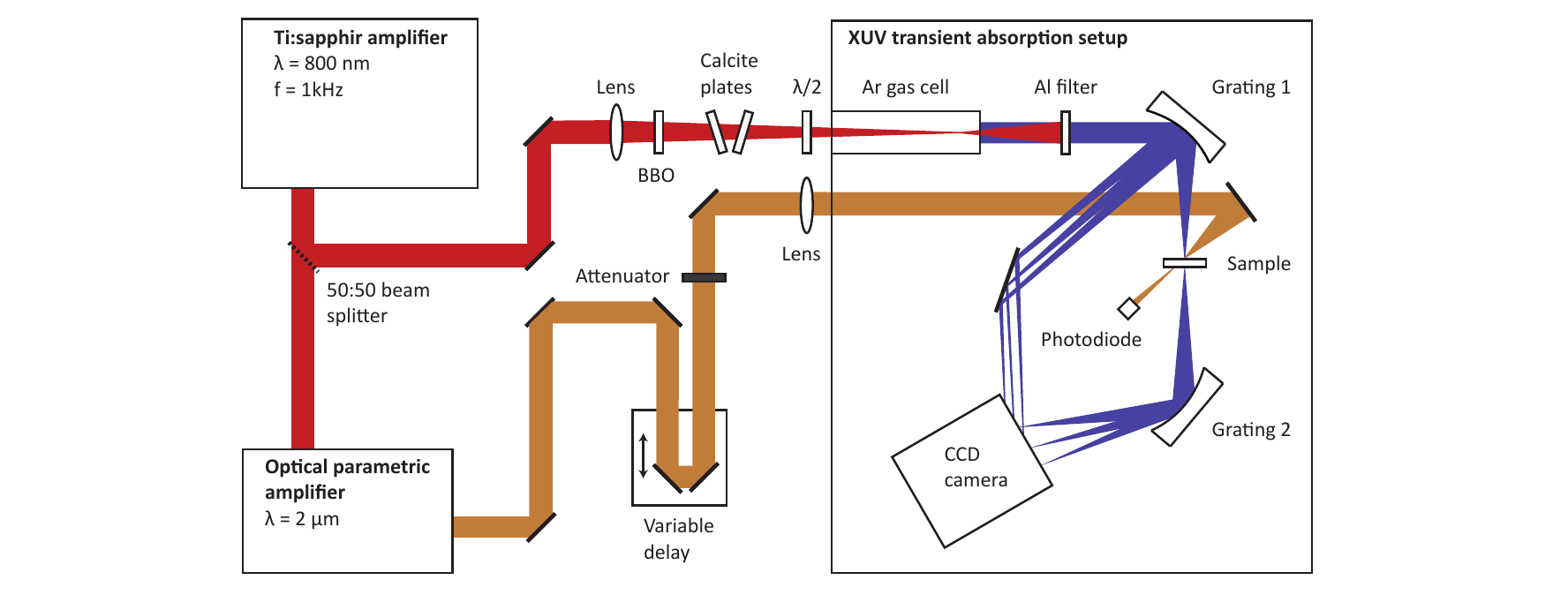}
\caption{Schematic of the transient absorption setup.}
\label{Fig:SetupDetail}
\end{figure*}

Infrared-pump XUV-probe core-level transient absorption measurements are carried out with a tabletop high-harmonic source (Fig.~\ref{fig:Setup}(a) and Fig.~\ref{Fig:SetupDetail}). It is driven by a $1\, \text{kHz}$ Ti:sapphire 35-femtosecond laser amplifier with a central wavelength of $800 \, \text{nm}$ and generates XVU radiation with spectral range of 25-50 eV in an Ar-filled gas cell. The high harmonic spectrum comprises peaks with $1.55 \, \text{eV}$ spacing, achieved by a bi-color laser excitation scheme \cite{kfirInlineProductionBicircular2016}. It consists of a BBO crystal to generate the second harmonic and two calcite plates to adjust the temporal overlap with respect to the fundamental pulse. The sample is excited by a $2 \, \mu \text{m}$ laser pulse with 40 fs pulse duration and an incidence angle of $5^{\circ}$ generated in an optical parametric amplifier (OPA). The wavelength was chosen to achieve a homogeneous absorption profile in the depth of the sample. Our setup consists of two toroidal gratings that spectrally disperse the XUV beam before the sample and after transmission through the sample. The two spectra, referred to as the reference and signal spectrum, are simultaneously detected on a charge-coupled device (CCD) camera. In combination with feed-forward neural network fitting \cite{heinrichArxiv}, this procedure achieves a sensitivity of $< 10^{-4}$ OD. Pump-probe traces are recorded with alternating pumped and un-pumped frames and randomly distributed timing delays, to avoid systematic drifts in time. 

Spatial overlap of pump and probe beams was achieved by scanning the sample in the focal plane and obtaining the position of the pump and probe beams through maps of the transmitted intensity. Two example transmission maps are shown in Fig.~\ref{Fig:Sample}(b) \& Fig.~\ref{Fig:Sample}(c). The integrated CCD counts and photocurrent of an additional infrared photodiode placed in the transmitted pump beam path measure the transmittance. In addition, these scans can be used for knife-edge measurements to estimate the probe and pump beam sizes on the sample as shown in Fig.~\ref{Fig:Sample}(e) and Fig.~\ref{Fig:Sample}(f). The spot profiles are analyzed by Gaussian fits to estimate spot diameters of $d_{\text{pump}} = 113 \, \mu \text{m}$ and $d_{\text{probe}} = 15 \, \mu \text{m}$ for pump and probe beam at full-width-of-half-maximum (FWHM), respectively. These dimensions are chosen such that the probe can be positioned well within the $100 \, \mu \text{m}$ windows while ensuring homogeneous excitation by the pump beam. 

To estimate the temporal resolution, the pump and probe pulse durations and the wavefront tilt originating from the non-collinear excitation are considered. By utilizing two-photon photo-absorption in a conventional silicon diode, intensity autocorrelation was performed on the $2 \, \mu \text{m}$ pump pulse. An autocorrelation width of $\tau_{\text{a.c.}} = 56 \, \text{fs}$ was extracted by fitting a Gaussian pulse shape as shown in Fig.~\ref{Fig:Sample}(d). The resulting pulse length of $\tau_{\text{pump}}=\tau_{\text{a.c.}}/\sqrt{2} = 40 \, \text{fs}$ is significantly larger than the expected $\tau_{\text{XUV}} \approx 10 \, \text{fs}$ pulse length of the XUV beam \cite{igarashiPulseCompressionPhasematched2012}. An additional contribution stems from the $5^{\circ}$ pulse front tilt between pump and probe pulses. This additional temporal shear over the probed length of $d = 15 \, \mu\text{m}$ can be computed by $\tau_{\text{tilt}}=d_{\text{probe}} \times \sin(5^{\circ})/c$ (c: speed of light) and is found to be $4.5 \, \text{fs}$. The combined temporal resolution is governed by the pump pulse width and is estimated to be $41 \, \text{fs}$ by convolution. 

\section{Absorbed energy}
\label{Appendix Fluence}
With the assumption of a Gaussian spot profile and nearly collinear excitation, the incident fluence is calculated as
\begin{equation}
F = P/(f \times r_{\text{pump}}^2 \times \pi) \, .
\end{equation}
Here, the $1/e$ radius of the pump pulse $r_{\text{pump}} =d_{\text{pump}}/(2 \sqrt{\ln(2)})$, the laser repetition rate $f$, and the measured pump power P are used. Further, the absorbed fluence can be estimated by subtracting reflected and transmitted portions for samples of length $L$: 
\begin{equation}
F_{abs}=F \times (1-R) \times (1-e^{-L/\delta} \times (1-R)).
\end{equation} 
From the literature, a reflectivity of R = 60\% (Ref. \citep{baylissReflectivityJointDensity1985}) and an absorption length of $\delta = 41 \, \text{nm}$ (Ref. \citep{baylissReflectivityJointDensity1985}) are used for the $2 \, \mu \text{m}$ pump wavelength. To obtain an accurate measure of the excitation density the absorbed fluence is converted to deposited energy per Ti atom (normal state unit cell) by:  
\begin{equation}
E_{u.c.}=F_{abs}/(L \times \rho) \, ,
\end{equation}
with the atomic density $\rho = 1.536 \times 10^{22}\,  \text{u.c.}/\text{cm}^3$ (Ref. \citep{riekelStructureRefinementTiSe21976}). However, for reflection studies on bulk samples, the formalism needs to be adjusted. Specifically, the length $L$ is set to half of the probe attenuation depth, and a second reflection is omitted. Table~\ref{Tab:Fluence} compares the incident fluence and absorbed energy of this work to selected literature values.
\begin{table}[h!]
   \centering
   \begin{ruledtabular}
    \begin{tabular}{c c c}
         Dataset & Fluence $F$ ($\mu \text{J}/\text{cm}^2$)&  Energy $E_{u.c.}$ (meV)\\
         \hline 
        This work& $260$, $660$&$5.9$, $15.0 $ \\
        Hedeyat \textit{et al}. \citep{hedayatExcitonicLatticeContributions2019}& $62$& $4.0 $\\
    \end{tabular}
    \end{ruledtabular}
    \caption{Excitation fluences and deposited energy per unit cell normalized to the probed sample volume.}
    \label{Tab:Fluence}
\end{table}

A significant portion of the absorbed energy is initially stored in the coherently excited phonon modes ${A_{1g}}^*$ and ${A_{1g}}$. From the TDDFT calculation, the maximal elongation from the equilibrium position $x_{i}$ of the atoms $i=$Ti, Se can be estimated which allows to calculate the potential energy for a specific phonon mode in the spring model by 
\begin{equation}
E=0.5 \times \sum_i  \alpha_{i} \omega^2 x_{i}^2 M_i \, .
\end{equation}
Here, $\omega$ is the phonon oscillation frequency, $M_i$ the atomic mass of the atom and $\alpha_{i}$ is the number of moving atoms per normal phase unit cell. For the ${A_{1g}}$ optical mode, both Se atoms ($\alpha_{Se} = 2$, $\alpha_{\text{Ti}} = 0$) in the unit cell are displaced by $x_{\text{Se}}= 0.0075 \, \AA$. In the case of the amplitude mode ${A_{1g}}^*$, we use $\alpha_{\text{Se}} = \frac{6}{4}$, $\alpha_{\text{Ti}} = \frac{3}{4}$, $x_{\text{Se}} = 0.0084  \, \AA$ and $x_{\text{Ti}} = 0.0255 \, \AA$, since only a fraction of the atoms participate in the mode. This leads to an energy of $E= 0.65 \, \text{meV}/\text{u.c.}$ and $E= 0.8 \, \text{meV}/\text{u.c.}$, stored in the coherently excited amplitude of the ${A_{1g}}$ and the ${A_{1g}}^*$ mode, respectively. Incoherent excitations of the modes are not captured here, and the total stored energy might even be larger. For the calculation, the excitation is assumed to be homogeneous over the sample, such that each unit cell is equally excited.

\section{Fitting procedure}
\label{Appendix Fitting}
\begin{figure*}[h!]
\centering
\includegraphics[width=\textwidth]{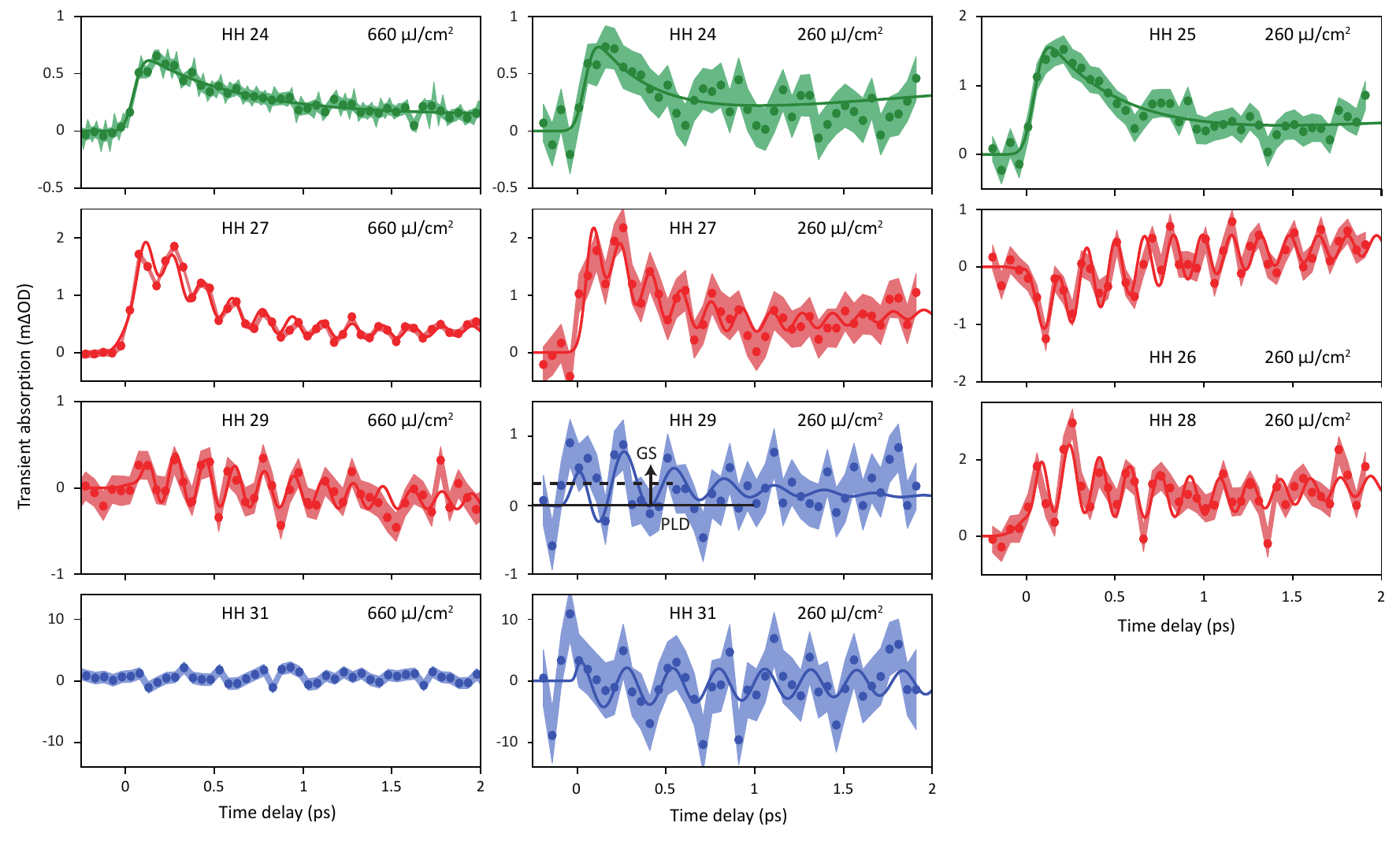}
\caption{Transient absorption traces of 1T-TiSe$_2$. Pump probe traces have been recorded for $F = 260 \, \mu \text{J}/\text{cm}^2$ and $F = 660 \, \mu \text{J}/\text{cm}^2$.}
\label{Fig:PPTraces}
\end{figure*}

To exctract phonon damping timescales, individual harmonics are fitted by incorporating all contributions: 
\begin{widetext}
\begin{align}
\label{eq:fitting}
    \Delta OD(\tau)= & ( \text{erf}\left(\frac{\tau}{T_{r1}}\right)+1)\times \left[A_e \times \text{exp}\left(- \frac{\tau}{T_e}\right)+A_b \times (1-\text{exp}\left(- \frac{\tau}{T_b}\right)) \right] + \\ \nonumber
    & (\text{erf} \left( \frac{\tau}{T_{r2}}\right)+1)\times  A_p \times \text{sin}(2 \pi  f \tau+\phi) \times \left[ \text{exp} \left(- \frac{\tau}{T_p} \right)+A_{e,p} \times \text{exp} \left(- \frac{\tau}{T_e} \right) \right]. 
\end{align}
\end{widetext}
Here, $T_{r1}$ and $T_{r2}$ represent the rise times after excitation, $A_e$, $A_b$ and $A_p$ the amplitude of the electron, phonon bath and coherent phonon contributions and $T_e$, $T_b$ and $T_p$ the corresponding damping times, respectively. One coherent phonon oscillation at frequency $f$ and its associated phase $\phi$ is considered. Since the time zero is not independently known, the phase is used as a time-zero offset and can not directly be used to extract a cosine behaviour indicative of displacive excitation. In addition to the direct contributions, a cross term between coherent phonon and electronic component needs to be included by an additional amplitude $A_{e,p}$. The fits are displayed in Fig.~\ref{Fig:Phonon}(a) and Fig.~\ref{Fig:PPTraces} and the fitting parameters for all harmonics are shown in Table~\ref{Tab:Fitting} for both fluences. 

\begin{table*}[h!]
    \centering
    \caption{Fitting parameters for individual high harmonics (HH). Shown are values for the 660 $\mu$J/cm$^2$ data set in the left and for the 260 $\mu$J/cm$^2$ data set in the right column.}
   \begin{ruledtabular}
    \begin{tabular}{ccccccccc} 
        HH & \multicolumn{2}{c}{$f \, \text{(THz)}$} & \multicolumn{2}{c}{$T_e \, \text{(ps)}$} & \multicolumn{2}{c}{$T_b \, \text{(ps)}$} & \multicolumn{2}{c}{$T_p \, \text{(ps)}$} \\ \hline
        25 & - & - & $0.8 \pm 0.1$ &$ 0.5  \pm0.2$ & $10  \pm3$ & $4  \pm6$ & - & - \\
        26 & $6.0  \pm0.1$& $6.1 \pm 0.1$ & $0.5 \pm 0.1$ & $0.2  \pm0.1$ & - & - & $1.9  \pm1.2$ & $1.3  \pm1.1$ \\
        27 & $6.0 \pm 0.1$& $5.8  \pm0.2 $& $0.5  \pm0.1$ & $0.4  \pm0.1$ & $6  \pm5$ & $5  \pm3$ & $1.5  \pm1.0$ & $0.8  \pm2.0$ \\
        28 & $6.0  \pm 0.1$&$  5.9  \pm0.1$&$ 0.5  \pm0.1$ & $0.5  \pm0.4$ & $2  \pm1$ & $4  \pm4$ & $1.1  \pm0.8$ & $1.4  \pm2.3$ \\
        29 & - & $3.5  \pm0.2$& - & - & - & - & - & $0.5  \pm2.6$
        \\
        30 & - & $3.4 \pm 0.2$ & - & - & - & - & - & $1.3  \pm1.2$\\
        31 & - & $3.4  \pm0.2$& - & - & - & - & - & $0.8  \pm0.8$ \\
    \end{tabular}
   \end{ruledtabular}
    
    \label{Tab:Fitting}
\end{table*}

\section{Simulations with Density Functional Theory (DFT)}
\label{Appendix DFT}
The Ti M$_{2,3}$ edge absorption spectra in Fig.~\ref{fig:Setup}(b), Fig.~\ref{fig:spectralShift}(b), and Fig.~\ref{Fig:Phonon}(d,e) are simulated with time-dependent density functional theory (TD-DFT) and the full-potential linearized augmented plane wave method (FP-LAPW) using the \textit{exciting} software package \cite{gulansExcitingFullpotentialAllelectron2014,vorwerkBetheSalpeterEquation2019,sagmeisterTimedependentDensityFunctional2009}. The calculations are performed with a Perdew-Burke-Ernzerhof (PBE) exchange-correlation functional \cite{perdewGeneralizedGradientApproximation1996}, and Local field effects are included in all TD-DFT calculations to incorporate the effects of many-body interactions on the core-level absorption \cite{aryasetiawanLocalfieldEffectsNiO1994,krasovskiiLocalFieldEffects1999,volkovAttosecondScreeningDynamics2019a}. Spectral calculations with atomic displacements along the A$_{1g}$ optical mode are conducted with a single unit cell on a $12\times 12\times 6$ $k$-grid. Equilibrium atomic positions and the displacements along the optical mode are obtained via geometry optimization with initial atomic positions taken from Ref.~\cite{perssonMaterialsDataTiSe22016} and subsequent phonon calculations at the $\Gamma$-point. Simulations of the core-level absorption spectra with displacements along the amplitude mode are conducted with a $2\times 2\times 2$ supercell. The atomic positions for the CDW and normal phase are taken from Ref.~\cite{biancoElectronicVibrationalProperties2015}. Here, the ground state DFT calculations are first carried out on a $12\times 12\times 6$ $k$-grid, and the following TD-DFT spectral simulations are conducted on a $6\times 6\times 3$ $k$-grid.

\begin{figure}[h!]
\centering
\includegraphics[width=\textwidth]{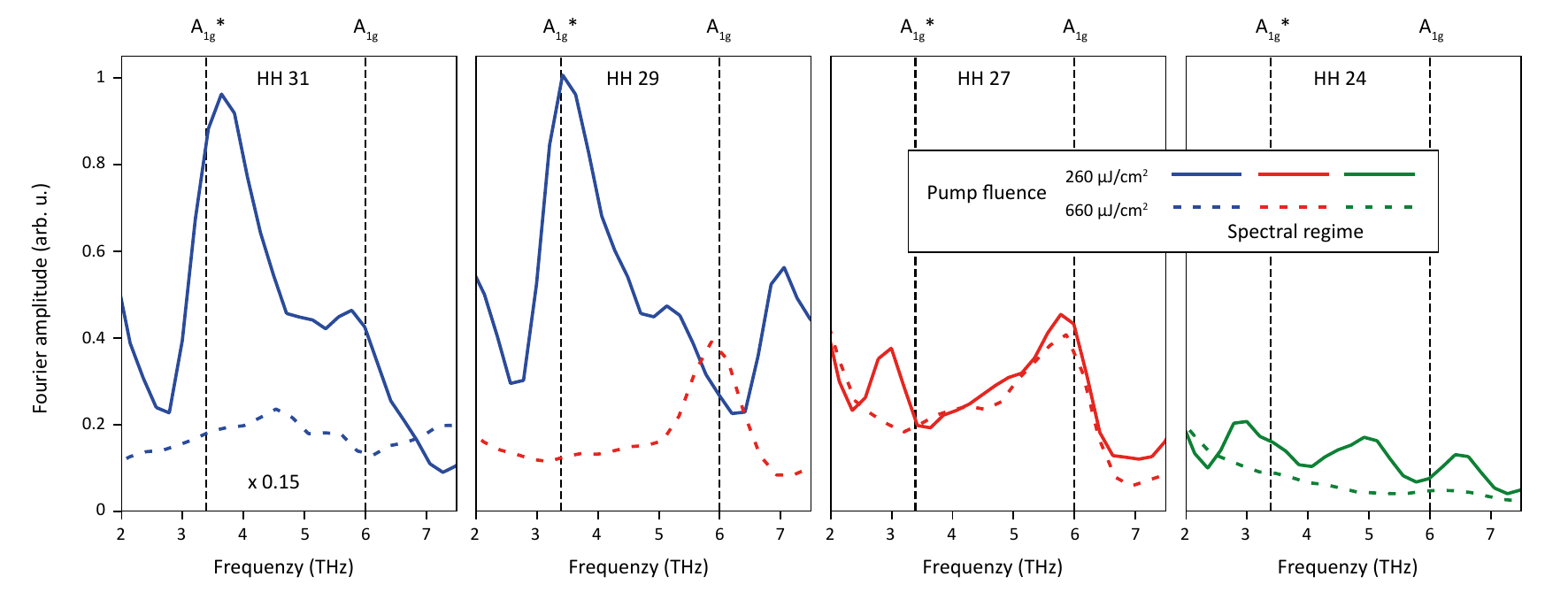}
\caption{Fourier spectra of pump probe traces.}
\label{Fig:Fourier}
\end{figure}

\bibliographystyle{apsrev4-2}

\end{document}